\documentclass[%
reprint,
superscriptaddress,
amsmath,amssymb,
aps,
pra,
]{revtex4-2}

\usepackage{graphicx}
\usepackage{dcolumn}
\usepackage{bm}
\usepackage{amssymb}
\usepackage{breqn}
\usepackage{xcolor}
\usepackage{amsmath}
\usepackage{float}
\usepackage{notoccite}

\begin{document}

\title{
Experimental study of decoherence of the two-mode squeezed vacuum state via second harmonic generation
}

\author{Fu Li}
\email{fuli@physics.tamu.edu}
\affiliation{
Institute for Quantum Science and Engineering$,$ Texas A\&M  University${,}$  College Station${,}$   TX  77843${,}$  USA
}
\affiliation{
Department of Physics and Astronomy${,}$  Texas  A\&M  University${,}$ College Station${,}$  TX  77843${,}$  USA
}

\author{Tian Li}
\email{tian.li@tamu.edu}
\homepage{T.L. and F.L. contributed equally to this work.}
\affiliation{
Institute for Quantum Science and Engineering$,$ Texas A\&M  University${,}$  College Station${,}$   TX  77843${,}$  USA
}
\affiliation{
Department of Biological and Agricultural Engineering${,}$  Texas A\&M  University${,}$  College Station${,}$ TX 77843${,}$  USA
}


\author{Girish S. Agarwal}
\affiliation{
Institute for Quantum Science and Engineering$,$ Texas A\&M  University${,}$  College Station${,}$   TX  77843${,}$  USA
}
\affiliation{
Department of Physics and Astronomy${,}$  Texas  A\&M  University${,}$ College Station${,}$  TX  77843${,}$  USA
}
\affiliation{
Department of Biological and Agricultural Engineering${,}$  Texas A\&M  University${,}$  College Station${,}$ TX 77843${,}$  USA
}

\date{\today}

\begin{abstract}

Decoherence remains one of the most serious challenges to the implementation of quantum technology. It appears as a result of the transformation over time of a quantum superposition state into a classical mixture due to the quantum system interacting with the environment. Since quantum systems are never completely isolated from their environment, decoherence therefore cannot be avoided in realistic situations. Decoherence has been extensively studied, mostly theoretically, because it has many important implications in quantum technology, such as in the fields of quantum information processing, quantum communication and quantum computation. Here we report a novel experimental scheme on the study of decoherence of a two-mode squeezed vacuum state via its second harmonic generation signal. Our scheme can directly extract the decoherence of the phase-sensitive quantum correlation $\langle \hat{a}\hat{b}\rangle$ between two entangled modes $a$ and $b$. Such a correlation is the most important characteristic of a two-mode squeezed state. More importantly, this is an experimental study on the decoherence effect of a squeezed vacuum state, which has been rarely investigated.


\end{abstract}


\pacs{42.50.Gy, 32.80.Qk, 42.65.-k}


\maketitle

\section{Introduction}

Realistic quantum systems are inevitably coupled with their environment. When a quantum system interacts with its environment, it will in general become entangled with a large number of environmental degrees of freedom~\cite{
ProgressInOptics,PhysRevA.64.063811,PhysRevLett.93.130406,PhysRevLett.94.043602,Boitier:2011ye,PhysRevA.98.020101,PhysRevLett.121.243402,PhysRevLett.123.140402}. It is this coupling between a quantum system and its environment that causes decoherence, sometimes also referred to as  environment-induced decoherence~\cite{Joos,Schlosshauer,PhysRevD.24.1516,PhysRevD.26.1862,RevModPhys.75.715,RevModPhys.76.1267}, 
which remains one of the most serious obstacles to the exploitation of quantum technology~\cite{Kwiat498,Almeida579,Yu598,Ladd,RevModPhys.90.035005}, although some protocols have been proposed~\cite{Esfahani_2016,doi:10.1098/rsta.2017.0315} and proof-of-principle experiment has been conducted~\cite{PhysRevLett.79.1964} on how the effect of decoherence can be reversed. Such coupling can be generally understood in terms of classical noise~\cite{PhysRevLett.97.140403},
such as in the investigations of optical parametric amplification~\cite{PhysRevA.47.3160,PhysRevLett.103.010501} and in the spectral diffusion theory that is widely used in, for instance, optical and magnetic resonance spectroscopy~\cite{LORING1985426,PhysRevA.32.2784}.
Stated in general terms, decoherence describes how interactions with the environment influence the statistics of results of future measurements on the quantum system. 

Decoherence happens all around us, and in this sense its consequences should be readily observed. There are several experimental areas that have played a key role in the experimental studies of decoherence: atom-photon interactions in a cavity~\cite{RevModPhys.73.565}, interferometry with mesoscopic molecules~\cite{Arndt_2005}, superconducting systems such as SQUIDs and Cooper-pair boxes~\cite{Leggett_2002}, and trapped ions~\cite{RevModPhys.75.281}. 
Recently, quantum nanomechanical systems also yield promising results for experimental tests of decoherence~\cite{RevModPhys.86.1391}. There are also some experimental investigations using decoherence for testing quantum mechanics~\cite{SCHLOSSHAUER20191}. These experiments are not only useful for evaluating the predictions of decoherence models, but also offering guidance for designing quantum devices that are capable of circumventing the detrimental influence of the environment.  

Among these prior experimental studies on decoherence, some are of particular interest to us due to the fact that they were conducted in an `all-optical' manner, for instance, Kwiat~\textit{et al.} used polarization entangled photon pairs produced by spontaneous parametric down-conversion (SPDC) to search for decoherence-free subspaces~\cite{Kwiat498}; Almeida~\textit{et al.} also used polarization entangled photon pairs to demonstrate that quantum entanglement may suddenly disappear although the environment-induced decay is asymptotic~\cite{Almeida579}. Both experiments employed the sophisticated quantum state tomography~\cite{10.1117/12.583241,10.1117/12.541178} to characterize the effect of decoherence. In this paper, we report a novel all-optical experimental scheme for studying the effect of decoherence. 
We showcase our ability to study the action of decoherence on a two-mode squeezed vacuum (TMSV) state in a gradual and controlled manner by measuring its second harmonic generation (SHG) signal from a Beta Barium Borate (BBO) crystal. The TMSV state is generated in the continuous-variable regime, hence a full density matrix tomography would not be applicable here. The decoherence is introduced by a neutral density (ND) filter to impose an uniform attenuation to the TMSV state.

\textcolor{black}{Our results are significant because the effect of decoherence on a squeezed vacuum state has been rarely investigated experimentally. More essentially, the most important property of a two-mode squeezed state is the nonzero quantum correlation $\langle \hat{a}\hat{b}\rangle$ between the two entangled modes $a$ and $b$, which ultimately determines the squeezing level of the state. Since direct measurement schemes such as intensity detections are not able to extract information about quantum correlation $\langle \hat{a}\hat{b}\rangle$, they can only extract information about correlation $\langle \hat{a}^\dagger \hat{a} \hat{b}^\dagger \hat{b}\rangle$, our experimental scheme therefore is set up in such a way that through the SHG signal induced by entangled photon pairs, information about the decoherence of the quantum correlation $\langle \hat{a}\hat{b}\rangle$ can be extracted explicitly.} 

\section{Theoretical Analysis}

\subsection {Cauchy-Schwartz inequity violation for the phase-sensitive correlation $\langle \hat{a} \hat{b} \rangle$}

It is well known that for quantum fields Cauchy-Schwartz inequalities (CSI) can be violated, because quantum fields can have $P$-distributions that do not have properties of a classical probability distribution. Such violations have been traditionally studied for intensity correlations like $\langle {\hat{a}^{\dagger2}} \hat{a}^2 \rangle$, $\langle \hat{a}^\dagger \hat{a} \hat{b}^\dagger \hat{b}\rangle$. Here we discuss a different correlation having phase-sensitive information. Let us first consider $a$ and $b$ to be complex random variables, then it is clear that for classical random variables, 

\begin{equation}
\begin{aligned}
\langle | c a + d b^\ast |^2 \rangle \geqslant	0 \ \ \ \ \ \forall \ c, d, 
\end{aligned}
\label{cCSI1}
\end{equation}
where $c$ and $d$ are arbitrary complex variables. This inequality for classical complex variables follows from the positivity of classical probability distributions. Equation~(\ref{cCSI1}) leads to 

\begin{equation}
\begin{aligned}
|c|^2 \langle |a|^2 \rangle + |d|^2 \langle |b|^2 \rangle + c^\ast d\langle a^\ast b^\ast \rangle + c d^\ast \langle a b \rangle  \geqslant 0, 
\end{aligned}
\label{cCSI2}
\end{equation}
and hence from the properties of the quadratic forms, it follows that 

\begin{equation}
\begin{vmatrix}
\langle |a|^2 \rangle & \langle a^\ast b^\ast \rangle \\
\langle a b \rangle & \langle |b|^2 \rangle  
\end{vmatrix} \geqslant	0. 
\label{cCSI3}
\end{equation}
The condition~(\ref{cCSI3}) leads to the CS inequality 

\begin{equation}
|\langle a b \rangle|^2  \leqslant \langle |a|^2 \rangle \langle |b|^2 \rangle.
\label{cCSI4}
\end{equation}

On the other hand, if we employ similar argument in the quantum domain using density matrices, then instead of Eq.~(\ref{cCSI1}), we can get 

\begin{equation}
\text{Tr}\{ \rho (c^\ast\hat{a}^\dagger+d^\ast\hat{b})(c\hat{a}+d\hat{b}^\dagger)\}\geqslant 0, 
\label{cCSI5}
\end{equation}
or

\begin{equation}
|c|^2\langle \hat{a}^\dagger \hat{a}\rangle + |d|^2\langle \hat{b}\hat{b}^\dagger\rangle + c^\ast d\langle \hat{a}^\dagger \hat{b}^\dagger\rangle + c d^\ast \langle \hat{a}\hat{b} \rangle \geqslant	0, 
\label{cCSI6}
\end{equation}
which leads to 

\begin{equation}
|\langle \hat{a}\hat{b} \rangle|^2 \leqslant \langle \hat{a}^\dagger \hat{a}\rangle \langle \hat{b} \hat{b}^\dagger \rangle = \langle \hat{a}^\dagger \hat{a}\rangle \langle \hat{b}^\dagger \hat{b} +1 \rangle. 
\label{cCSI7}
\end{equation}

On comparing Eq.~(\ref{cCSI7}) with Eq.~(\ref{cCSI4}), we can see that quantum fields with nonzero phase-sensitive correlations will always violate the classical inequity~(\ref{cCSI4}). All this discussion is based on the fact that quantum optical detectors measure the normally ordered correlations. \textcolor{black}{The Cauchy-Schwartz inequity violation for the phase-sensitive correlation $\langle \hat{a} \hat{b} \rangle$ verified here is to show that the correlation $\langle \hat{a} \hat{b} \rangle$ is nonclassical. It is this nonclassical correlation that is responsible for the quantum features of the two-mode squeezed state. Since the whole idea of our work is to extract where the nonclassical contribution is, therefore the experiment is set up in such a way that this nonclassical correlation $\langle \hat{a} \hat{b} \rangle$ and its decoherence can be extracted explicitly.}

\begin{figure*}[hbtp]
    \includegraphics[width=0.75\textwidth]{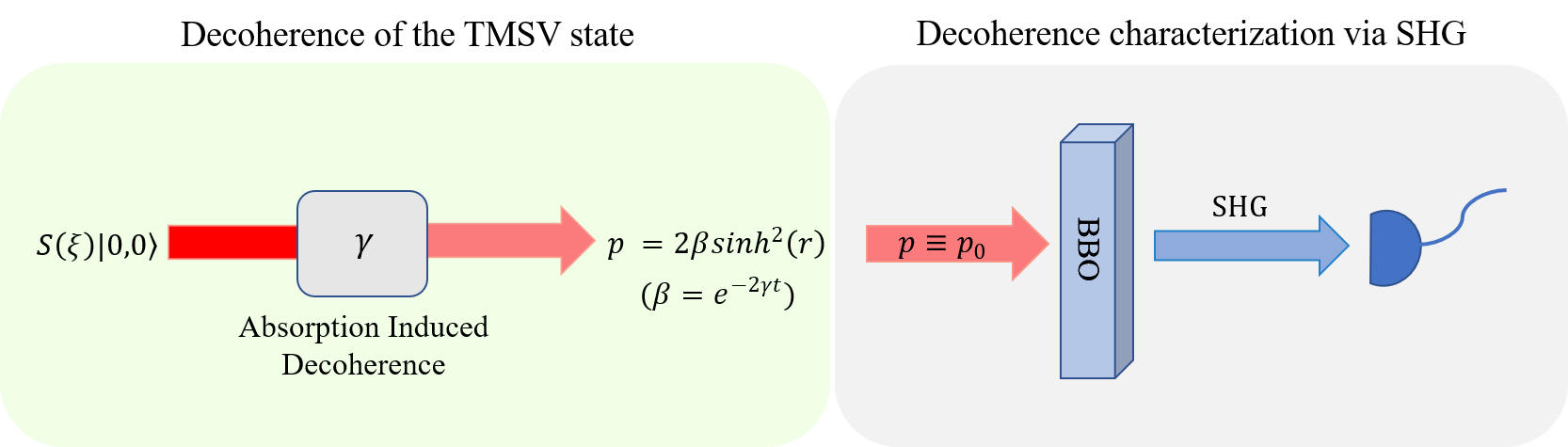}
    \caption{Theoretical decoherence characterization of the TMSV state via its SHG signal. To isolate the effect of decoherence on the quantum correlation $\langle \hat{a} \hat{b} \rangle$, experiment is performed by holding power $p$ constant, i.e., $p\equiv p_0$.}
    \label{Decoherence}
\end{figure*}

\subsection{Decoherence characteristics of the two-mode squeezed vacuum state via amplitude damping}

Let us consider a TMSV state given by $\hat{S}\left(\xi\right)\left|0,0\right\rangle$, where $\hat{S}\left(\xi\right)=e^{\xi \hat{a}^{\dagger}\hat{b}^{\dagger}-\xi^{\ast}\hat{a}\hat{b}}$ is the two-mode squeezing operator and $\xi=re^{i\theta}$, where $r$ is the squeezing parameter. The mean number of photons for each mode is the same, i.e., $\left\langle \hat{a}^{\dagger}\hat{a}\right\rangle =\left\langle \hat{b}^{\dagger}\hat{b}\right\rangle =\sinh^{2}r$, and we let $p_{0}$ represent the mean number of photons of the state, then
\begin{equation}
p_{0}=\left\langle \hat{n}\right\rangle =\left\langle \hat{a}^{\dagger}\hat{a}+\hat{b}^{\dagger}\hat{b}\right\rangle =2\sinh^{2}r. 
\label{power}
\end{equation}

Let $\rho$ be the density matrix of the TMSV state, $\rho=\hat{S}\left(\xi\right)\left|0,0\right\rangle \left\langle 0,0\right|\hat{S}^{\dagger}\left(\xi\right)$. The decoherence of the TMSV state is described by the master equation:
\begin{dmath}
\frac{\partial \rho}{\partial t}=-\gamma\left(\hat{a}^{\dagger}\hat{a}\rho-2\hat{a}\rho\hat{a}^{\dagger}+\rho\hat{a}^{\dagger}\hat{a}\right)-\gamma\left(\hat{b}^{\dagger}\hat{b}\rho-2\hat{b}\rho\hat{b}^{\dagger}+\rho\hat{b}^{\dagger}\hat{b}\right),
\label{DensityMatrix}
\end{dmath}
where $\gamma$ gives the decay of the field amplitude. The dynamical equation Eq.~(\ref{DensityMatrix}) is solved subject to the initial condition $\rho(t=0)=\hat{S}\left(\xi\right)\left|0,0\right\rangle \left\langle 0,0\right|\hat{S}^{\dagger}\left(\xi\right)$. Instead of presenting a time-dependent solution for the full density matrix, we present the result for the normally-ordered correlations of arbitrary order. It turns out that 
\begin{dmath}
\left\langle\left(\hat{a}^{\dagger}\right)^{m}\hat{a}^{n}\left(\hat{b}^{\dagger}\right)^{p}\hat{b}^{q}\right\rangle_{t}=\beta^{\left(m+n+p+q\right)/2}\left\langle \left(\hat{a}^{\dagger}\right)^{m}\hat{a}^{n}\left(\hat{b}^{\dagger}\right)^{p}\hat{b}^{q}\right\rangle _{0}, 
\label{DensityMatrix3}
\end{dmath}
where $\beta=e^{-2\gamma t}$ and $1-\beta$ represents the absorption of the modes. The subscript `$0$' denotes $t=0$. In particular, 
\begin{equation}
\begin{aligned}
\langle \hat{a}^\dagger \hat{a} \rangle_t = \beta \langle \hat{a}^\dagger \hat{a} \rangle_{0}, \ \langle \hat{b}^\dagger \hat{b} \rangle_t = \beta \langle \hat{b}^\dagger \hat{b} \rangle_{0},\\
\langle \hat{a}^\dagger \hat{a} \hat{b}^\dagger \hat{b}\rangle_t = \beta^2 \langle \hat{a}^\dagger \hat{a} \hat{b}^\dagger \hat{b}\rangle_{0}. 
\label{IO}
\end{aligned}
\end{equation}
The correlation $\langle \hat{a}^\dagger \hat{a} \hat{b}^\dagger \hat{b}\rangle$ for the TMSV state is well known to be given by
\begin{equation}
\begin{aligned}
\langle \hat{a}^\dagger \hat{a} \hat{b}^\dagger \hat{b}\rangle=\langle \hat{a}^\dagger \hat{a} \rangle \langle \hat{b}^\dagger \hat{b}\rangle + |\langle \hat{a}\hat{b} \rangle|^2 \\ 
=\sinh^{2}r\times\cosh^{2}r+\sinh^{4}r,
\end{aligned}
\end{equation}
the phase-sensitive correlation $|\langle \hat{a}\hat{b} \rangle|$ has the value of $\sinh r\times\cosh r$ and satisfies the equality sign in Eq.~(\ref{cCSI7}). The decoherence of the intensity-intensity correlation $G_{ab}^{(2)}$ between the two modes is therefore given by 
\begin{equation}
\begin{aligned}
G_{ab}^{(2)}=\langle \hat{a}^\dagger \hat{a} \hat{b}^\dagger \hat{b}\rangle_t=\beta^2(\sinh^{2}r\times\cosh^{2}r+\sinh^{4}r).
\label{G2}
\end{aligned}
\end{equation}

\textcolor{black}{It should be noted that, all the discussion above is based upon the assumption that the generated TMSV state is a pure state. In general, the state purity can be affected by the loss in the atomic medium. In order to avoid this problem, our experiment is performed under conditions that the intermediate state detuning is larger than the half Doppler width and the two-photon detuning is much bigger than the ground state decoherence. Under these conditions, the generated TMSV state is for all practical purposes a pure state. \textcolor{black}{It is worth mentioning that, according to the simulation reported in Ref.~\cite{PhysRevA.78.043816}, in the range of gain of less than 10 (our maximal gain is approximately 8, see Fig.~\ref{ConevsPump} below), the total loss on the probe beam due to atomic absorption of probe photons would be less than 10 \% (see Fig.~5 in Ref.~\cite{PhysRevA.78.043816}), and the absorption of conjugate photons would be negligible since it is far detuned from the atomic resonance. With this amount of loss sustained by the probe beam \textit{in the source}, we show in the Appendix that the generated TMSV state would be indeed still close to a pure state.} We also would like to emphasize that the quantum correlation $\langle \hat{a}\hat{b}\rangle$ is extracted from the study of intensity correlations obtained from the SHG signal's mean power level, which is determined by the quantum statistics of the input field.} 

In the next section we outline our procedure for studying the decoherence of the quantum correlation $\langle \hat{a}\hat{b} \rangle$.

\subsection{Characterization of decoherence of the two-mode squeezed vacuum state via its SHG signal}

In this section, we demonstrate that the decoherence of a TMSV state can be characterized via its SHG signal from a BBO crystal. The SHG signal is proportional to the intensity-intensity correlation $G_{ab}^{(2)}$ given by Eq.~(\ref{G2}). We first note that the squeezing parameter $r$ is proportional to the power $P_0$ of the pump light that is used to produce the TMSV state via the four-wave mixing (FWM) process (see details in Section III). As shown in Eq.~(\ref{IO}), the output power
\begin{equation}
\begin{aligned}
p_{\text{out}}=2\beta\sinh^{2}r, \\ r \equiv \alpha P_0, 
\label{outpower}
\end{aligned}
\end{equation}
where $\alpha$ is related to the strength of the FWM process. 

While the absorber attenuates the power of the TMSV state (see Fig.~\ref{Decoherence}), we make up for the loss of power by increasing the power of the pump, namely,
\begin{equation}
\begin{aligned}
2\beta\sinh^{2}r=2\beta\sinh^{2}(\alpha P_0)=2\sinh^{2}(\alpha P'_0)\equiv p_0,
\label{outpower1}
\end{aligned}
\end{equation}
where $p_0$ is the fixed power value. Clearly $P'_0$ depends on the parameter $\beta$. 

We perform the SHG measurements by changing the absorption (varying $\beta$) but holding constant the number of photons in the TMSV beam emerging from the decohering mechanism, i.e., the absorber. Thus when $\beta$ is changed, then $P'_0$ is changed appropriately as determined by Eq.~(\ref{outpower1}). Therefore, by holding $p_0$ constant, the intensity-intensity correlation $G_{ab}^{(2)}$ in Eq.~(\ref{G2}) can be rewritten as
\begin{equation}
\begin{aligned}
G_{ab}^{(2)} =\frac{p_{0}^{2}}{2}+\frac{p_{0}}{2}\times\beta. 
\label{G2_1}
\end{aligned}
\end{equation}
It is important to note that the occurrence of the $\beta$ term in Eq.~(\ref{G2_1}) can be traced back to the presence of the unity in Eq.~(\ref{cCSI7}).

In the experiment we use a ND filter to impose decoherence, and the transmission coefficient 
$\beta=10^{\text{-ND}}$. Then the relation between the SHG signal induced by the \textit{partially-decoherent} TMSV state and the attenuation ND can be therefore readily obtained: 

\begin{equation}
\begin{aligned}
\text{SHG} \propto G_{ab}^{(2)}=\frac{p_{0}^{2}}{2}+\frac{p_{0}}{2}\times10^{\text{-ND}}.
\end{aligned}
\label{fit}
\end{equation}

 It should be noted that for a classical beam of light with power $p_{0}$, its SHG signal is $\propto p_{0}^{2}$, thus the linear term $(p_{0}/2)\times 10^{\text{-ND}}$ in Eq.~(\ref{fit}) is solely due to the quantum property of the correlation $\langle \hat{a} \hat{b} \rangle$, and it can only be degraded but never vanishes just by imposing absorption to the TMSV state. Another interesting aspect that emerges from Eq.~(\ref{fit}) is that because of the way we set up the experiment, the decoherence term shows up in the \textit{linear} regime of the power dependence of the SHG signal.
                                                
\textcolor{black}{It is also worth mentioning that a related but different experiment has been conducted by Dayan \textit{et al.} in Ref.~\cite{PhysRevLett.94.043602}, where they experimentally demonstrated sum-frequency generation (SFG) with entangled photon pairs produced with SPDC. Their focus is on the sum-frequency generation due to quantum light from SPDC. Basically, entangled photon pairs down-converted in one crystal are up-converted in the second one to produce the SFG photons. They showed that by changing the pump power, thus \textit{changing the power of the entangled photon pairs at the second crystal}, the SFG rate behaves in a close-to-linear manner. In our work, however, we \textit{keep the power at the BBO crystal constant} but use ND filters to effectively change the quantum correlation $\langle \hat{a} \hat{b} \rangle$ between the two entangled modes $a$ and $b$ of the TMSV state. This is the major difference and the focus of our study. Besides we worked with entangled photon pairs produced with the four-wave mixing (FWM) process whereas they worked with SPDC.}

\section{Experiment and Results}

\subsection{Experimental setup}

\begin{figure}[hbtp]
    \begin{center}
    \includegraphics[width=1.00\linewidth]{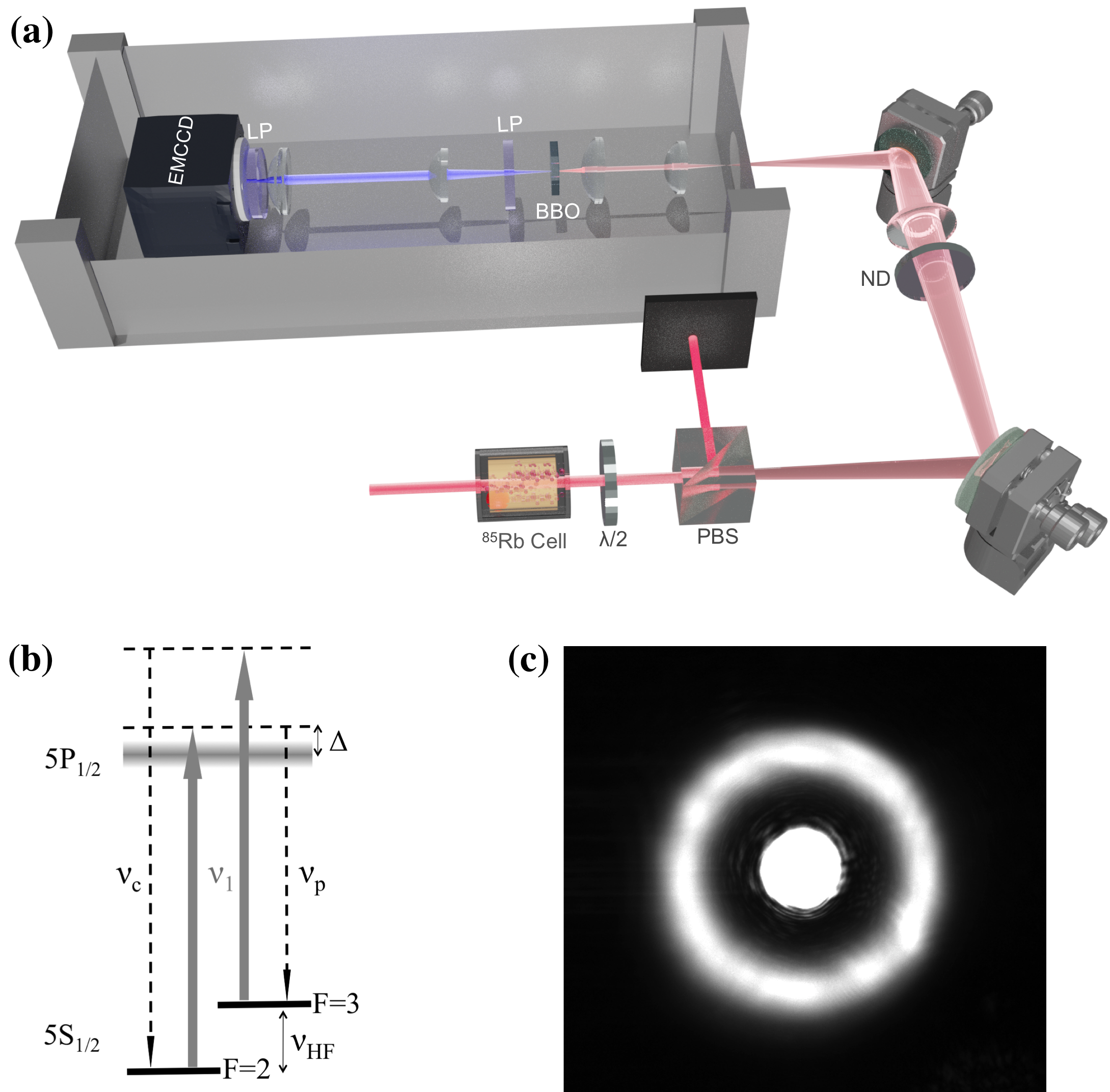}
    \caption{
     (a) Experimental setup in which a CW laser-pumped $^{85}$Rb vapor cell produces a TMSV state via the FWM process. The TMSV beam (i.e. the `light cone') is separated from the pump beam by a $\sim$~$2\times10^5$~$:1$ polarizing beam splitter after the cell. The SHG signal from the BBO crystal is collected by an EMCCD camera. Two low-pass filters are mounted in front of the camera to eliminate undesired excitation photons. The BBO crystal and the EMCCD camera are enclosed in a light-proofing box to block ambient light. PBS: polarizing beam splitter, ND: neutral density filter, LP: low-pass filter. (b) Level structure of the D1 transition of $^{85}$Rb atom. The optical transitions are arranged in a double-$\Lambda$ configuration, where $\nu_p$, $\nu_c$ and $\nu_1$ stand for probe, conjugate and pump frequencies, respectively, fulfilling $\nu_p$ +  $\nu_c$ =  $2\nu_1$ and $\nu_c - \nu_p = 2\nu_{HF}$. The width of the excited state in the level diagram represents the Doppler broadened line. $\Delta$ is the one-photon detuning. $\nu_{\text{HF}}$ is the hyperfine splitting in the electronic ground state of $^{85}$Rb. (c) Image of the cross-section of the TMSV state, i.e., the `light cone', where the bright central spot is the residual pump beam. \textcolor{black}{The angular separation between the center of the ring and the center of the residual pump beam is $\sim 0.3^{\circ}$, and the width of the ring is measured to be $\sim 0.05^{\circ}$.}
        \label{Setup}}
    \end{center}
\end{figure}

The setup of our experimental scheme is shown in Fig.~\ref{Setup}(a). We generate the TMSV state with the FWM process in a $^{85}$Rb atomic vapor cell. The vapor cell is kept at 117~$^{\circ}$C to maintain enough atom number density for the interaction. The respective $^{85}$Rb atomic level structure is shown in Fig.~\ref{Setup}(b). The atomic medium is pumped by a strong (up to $\sim 1.2$~W) narrow-band continuous-wave (CW) laser at frequency $\nu_1$ ($\lambda = 795$~nm) with a typical linewidth $\Delta \nu_1 \sim 100$~kHz and 700~$\mu$m $1/e^2$ radius. The pump laser is blue-tuned by a \textcolor{black}{`one-photon detuning $\Delta$' of 900~MHz} with respect to the $^{85}$Rb $5S_{1/2}, F= 2 \rightarrow 5P_{1/2}$, D1 transition.  Due to the FWM parametric process, two pump photons are converted into a pair of twin photons, namely `probe $\nu_p$' and `conjugate $\nu_c$' photons, adhering to the energy conservation $2 \nu_1 = \nu_p + \nu_c$ (see the level structure in Fig.~\ref{Setup}(b)). These twin photons are separated in frequency by twice of the hyperfine splitting in the electronic ground state of $^{85}$Rb, i.e., $\nu_c - \nu_p = 2\nu_{HF}$. The finite length of the atomic vapor cell (12.5~mm) slightly relaxes the longitudinal phase matching condition and allows for a range of angles, which effectively sets the angular acceptance of the FWM process, and produces the TMSV state in a form of `light cone' after the cell. The cross-section of the cone is shown in Fig.~\ref{Setup}(c), where the bright central spot is the residual pump beam not completely filtered out by the PBS at the exit of the cell (See Fig.~\ref{Setup}(a)). \textcolor{black}{The angular separation between the center of the ring and the center of the residual pump beam is $\sim 0.3^{\circ}$, and the width of the ring is measured to be $\sim 0.05^{\circ}$.} In the experiment, the central spot is blocked by a small black metal disk mounted on an extremely thin wire, so that the integrity of the TMSV state is not affected. \textcolor{black}{The power of the TMSV state is measured by focusing this light cone on to a power meter sensor composed of a silicon photodiode.}

After the cell, the TMSV state is collimated to $\sim$~1~mm $1/e^2$ radius before focused on a BBO crystal by a 16~mm aspheric lens. The SHG signal is collected by an electron-multiplying charge-coupled-device (EMCCD) camera, in front of which two low-pass filters are mounted to eliminate undesired excitation photons. \textcolor{black}{A region in the image captured by the EMCCD camera that contains the whole beam profile of the SHG signal is cropped, and photon counts in each pixel are added together to obtain the total photon counts/intensity of the SHG signal during an integration time of \textcolor{black}{1~s}. The use of an EMCCD camera for photon-counting measurement has become a common practice~\cite{Jechow:2013rc,Upton:2013fk} given that it is much easier to be implemented than single photon counting modules, which typically would require fiber coupling.} The decoherence measurement stage including the BBO crystal and the EMCCD camera are enclosed in a light-proofing box to block ambient light. A ND filter is mounted in front of the box to introduce decoherence to the TMSV state.

\subsection{Results}

\begin{figure}[hbtp]
    \begin{center}
    \includegraphics[width=1.00\linewidth]{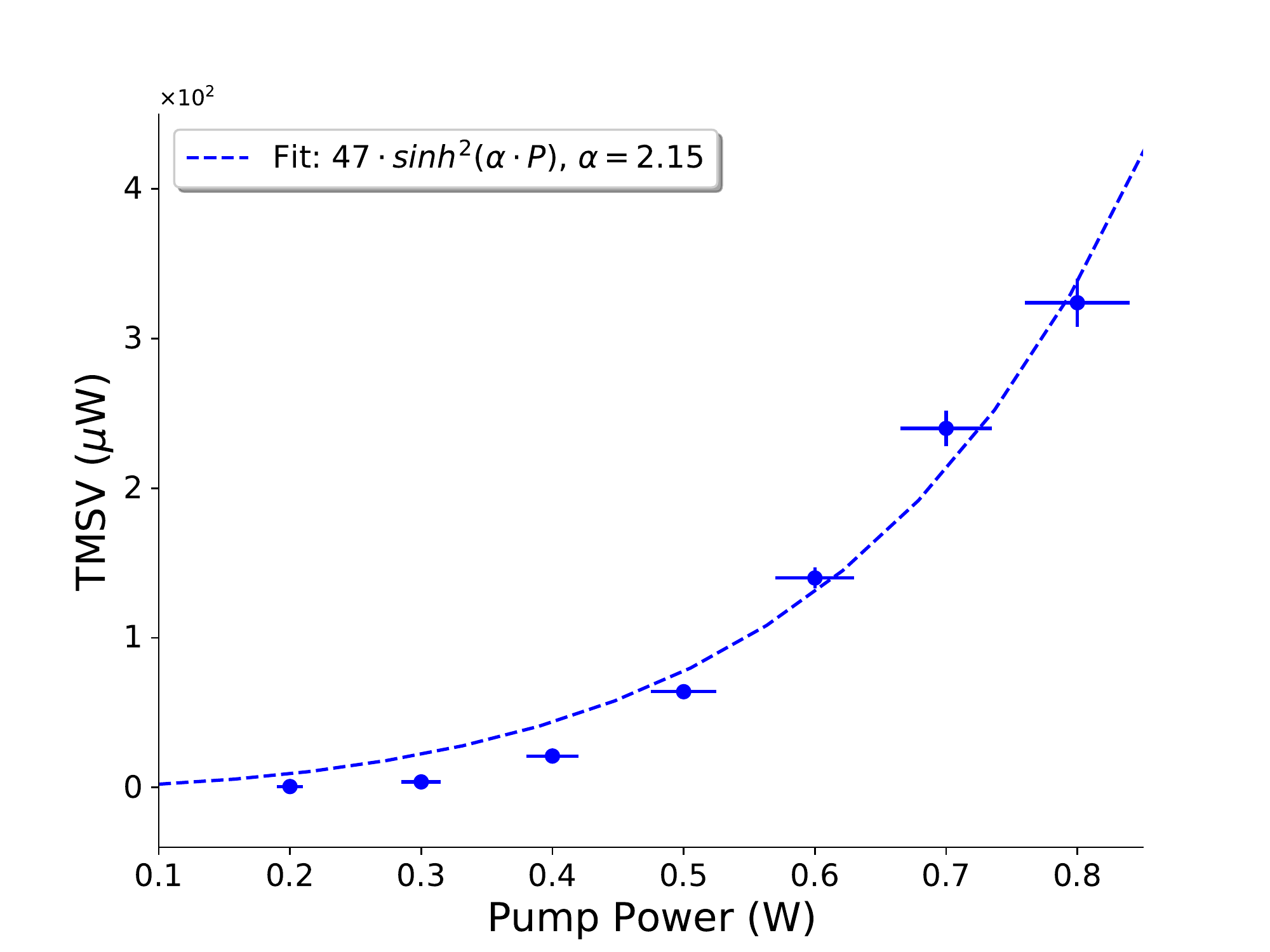}
    \caption{TMSV beam (light cone) power as a function of pump power. Dashed line is the theoretical fit according to Eq.~(\ref{power}) with $r \equiv \alpha P$, where $P$ is the pump power. 
    \label{ConevsPump}}
    \end{center}
\end{figure}

We first characterize the TMSV beam produced with the FWM process in the $^{85}$Rb vapor cell. As we increase the pump power $P$, the TMSV beam (light cone) power $p_{\text{TMSV}}$ should also increase according to Eq.~(\ref{power}) since the squeezing parameter $r$ is linearly proportional to the pump power $P$.  In Fig.~\ref{ConevsPump} we plot the data along with the theoretical fit, $p_{\text{TMSV}}\propto\sinh^2({\alpha P})$, where $\alpha$ is the proportional constant. \textcolor{black}{Note that the light cone (Fig.~\ref{Setup}(c)) contains very large number of independent TMSV modes, including both temporal and spatial modes. Therefore the power of the light cone would be the summation over all individual modes, resulting in power in the micro-Watt range in Fig.~\ref{ConevsPump}.} The error bars on both axes represent the statistical uncertainties of one standard deviation together with the 5~\% measurement uncertainties introduced by the power meters. \textcolor{black}{Note that although our pump power can be increased up to 1.2~W, this fit is obtained by only using data points with pump powers less than 0.8~W as we suspect the gain saturation effect might start setting in when the pump power is greater than 0.8~W. By keeping data points with lower powers, we calculate that our maximal gain is approximately 8 with the used maximal pump power $\text{P}_{\text{max}} = 0.8$~W, i.e., $\text{G}_{\text{max}} = \text{cosh}^2  r_{\text{max}}  = \text{cosh}^2(\alpha \cdot \text{P}_{\text{max}}) = \text{cosh}^2 (2.15\times0.8)  \cong 8$. According to Ref.~\cite{PhysRevA.78.043816}, with a gain of less than 10 the total loss sustained by the probe beam \textit{in the cell} would be less than 10~\%, and the loss sustained by the conjugate beam \textit{in the cell} would be negligible. It is therefore safe to assume that the data points in Fig.~\ref{ConevsPump} are all in the pure states as their pump powers are less than 0.8~W, i.e., their gains are less than 8 (see the Appendix for a detailed verification).}

\begin{figure}[hbtp]
    \begin{center}
    \includegraphics[width=1.00\linewidth]{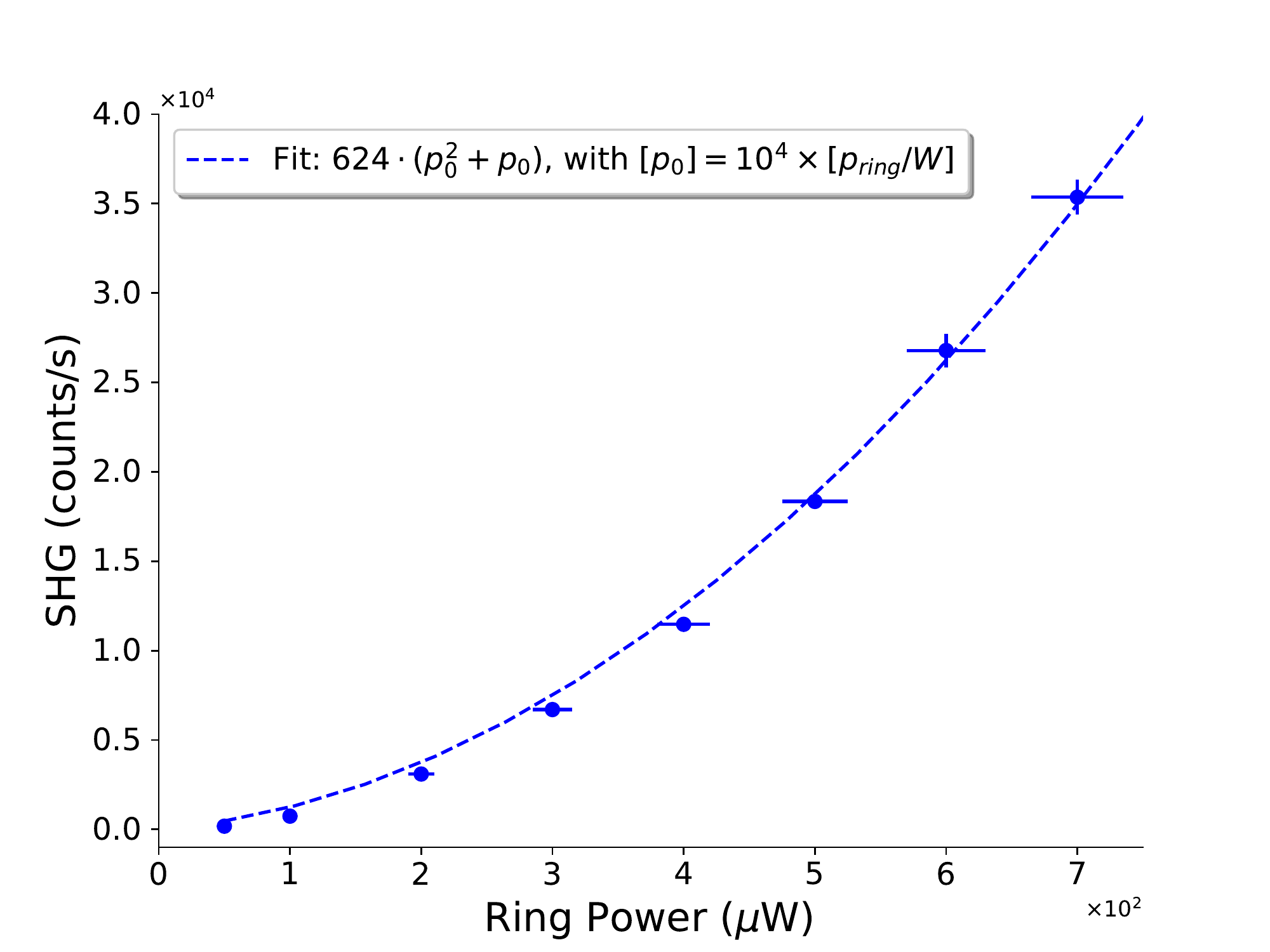}
    \caption{
    SHG signal as a function of the power of the light cone (shown in Fig.~\ref{Setup}(c) as the `ring'). Dashed line is a polynomial fit according to Eq.~(\ref{fit}) with $\text{ND} = 0$. 
        \label{SHGvsCone}}
    \end{center}
\end{figure}

We also note that according to Eq.~(\ref{fit}), when there is no ND filter ($\text{ND} = 0$) the SHG signal follows a polynomial dependence on the TMSV state power $p_0$, i.e., $\text{SHG} \propto p_0^2/2+p_0/2$. In order to verify this polynomial functionality, the measured SHG signals versus $p_0$ are plotted in Fig.~\ref{SHGvsCone} as blue dots. They agree very well with a polynomial behavior, represented by the fit function $624\times(p^2_0+p_0)$. Note that here $p_0$ has the dimension of $10^4 \times [p_{\text{ring}}/\text{W}$], where $p_{\text{ring}}$ is the power of the light cone. 

\begin{figure}[hbtp]
    \begin{center}
    \includegraphics[width=1.00\linewidth]{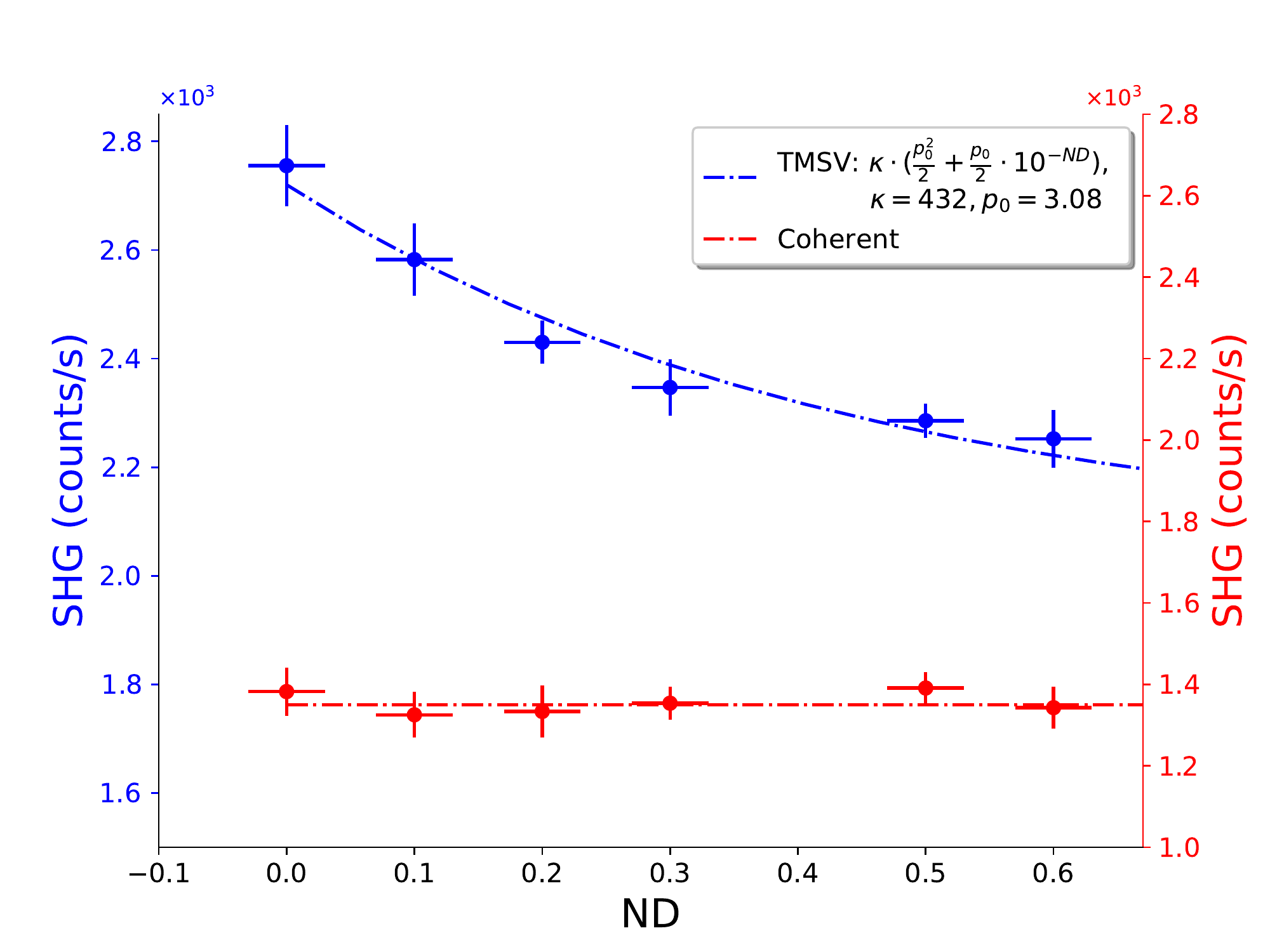}
    \caption{
    SHG signals induced by 80~$\mu$W partially-decoherent TMSV beams of light (blue dots, left-hand side $y$-axis) and $80~\mu$W coherent beams of light (red dots, right-hand side $y$-axis) as a function of transmission of ND filters. Dashed blue line is the theoretical fit according to Eq.~(\ref{fit}). 
        \label{Decoh}}
    \end{center}
\end{figure}

In order to study the effect of decoherence on the TMSV state imposed by an uniform attenuation, the optical density of the ND filter, i.e., the absorption level, is increased while the input beam (i.e., the `partially-decoherent' TMSV beam) power before the BBO crystal is fixed at 80~$\mu$W. This can be done by increasing the pump power appropriately (see Fig.~\ref{ConevsPump}). In Fig.~\ref{Decoh}, we plot the SHG signal induced by the partially-decoherent TMSV beam as a function of the transmission of the ND filter. Blue dots are the SHG measurements for $80~\mu$W partially-decoheret TMSV beams of light, plotted with $y$-axis on the left-hand side of the graph. The dashed blue line represents the fit function $(\kappa/2)\times (p_{0}^{2}+p_{0}\times10^{\text{-ND}})$ according to Eq.~(\ref{fit}) with fitting parameters $\kappa = 432$ and $p_{0} = 3.08$. This theoretical fit yields an excellent agreement with experimental observations. \textcolor{black}{It can be seen from Fig.~\ref{ConevsPump} that the TMSV state of $80~\mu$W is generated with pump power of 0.5~W, which implies a squeezing parameter $r$ of 1.1 ($r=\alpha P=2.15\times0.5\cong1.1$). According to Eq.~(\ref{power}) the mean number of photons in this TMSV state can be calculated as $p_0 = 2\text{sinh}^2 r = 2\times \text{sinh}^2 (1.1) = 3.57$, which is greater than 3.08 obtained from the fit in Fig.~\ref{Decoh}. This discrepancy can be attributed to the apparent fact that not all the photons in the light cone (shown in Fig.~\ref{Setup}(c) as the `ring') can be involved in the SHG process in the BBO crystal due to the phase matching condition.
}As a comparison, we also plot SHG signals induced by $80~\mu$W coherent beams of light with $y$-axis on the right-hand side of the graph. As shown in Fig.~\ref{Decoh} with these red dots and as what we expected, they form a flat line due to the fact that statistical properties of coherent light are indifferent to absorption.

Notice that it may not be fair to compare the `absolute' SHG signals induced by the TMSV beam and the coherent beam in Fig.~\ref{Decoh} because we were not able to make the coherent beam a `doughnut' shape like the TMSV beam, therefore they had different phase-matching conditions at the BBO crystal. Also note that, the first data point in Fig.~\ref{Decoh} is the SHG signal induced by $80~\mu$W TMSV beam without a ND filter, which is slightly higher than the second data point in Fig.~\ref{SHGvsCone} induced by $100~\mu$W TMSV beam. We attribute this discrepancy again to different phase-matching conditions (due to slightly different optical alignments) under which these two sets of measurements were taken.

\section{Conclusions}

We demonstrate a novel and unsophisticated all-optical experimental scheme for studying the decoherence effect on a TMSV state. The TMSV state is generated with the FWM process in an atomic $^{85}$Rb vapor cell, and the decoherence is characterized through the SHG signal induced by the TMSV state from a BBO crystal. Although squeezed state nowadays has become an extremely versatile tool for precision measurements and for interferometry due to its capability of offering unprecedented measurement sensitivity~\cite{RevModPhys.90.035005,doi:10.1063/5.0010909}, the decoherence effect of a squeezed state has rarely been investigated experimentally. Our scheme therefore would make a great addition to the research on decoherence of nonclassical states. The significance of our experiment resides in the fact that it demonstrates our capability of directly extracting the decoherence of quantum correlation $\langle \hat{a}\hat{b}\rangle$ between two entangled modes $a$ and $b$, which is the most important property of a two-mode squeezed state. It also showcases the possibility of characterizing the effect of decoherence in a controllable and measurable manner on a quantum state in the continuous-variable regime. 

\section*{Acknowledgements}

We gratefully acknowledge the support of Air Force Office of Scientific Research (Award No. FA-9550-20-1-0366) and the Robert A. Welch Foundation (Grant No. A-1943). F.L. acknowledges support from the Herman F. Heep and Minnie Belle Heep Texas A\&M University Endowed Fund administered by the Texas A\&M Foundation. 


\textcolor{black}{\section*{Appendix}}

\textcolor{black}{\subsection*{Pure state analysis validation}}

\begin{figure*}[tbhp]
    \includegraphics[width=\linewidth]{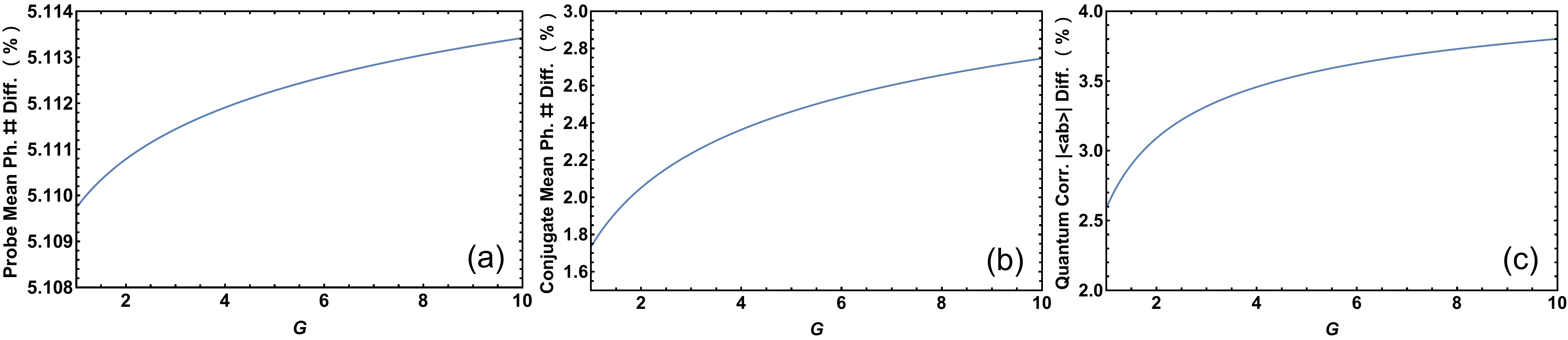}
    \caption{
    The mean number of photons difference in percentile at the exit of the cell between $\alpha_a = 0.1$ and $\alpha_a = 0$ versus the FWM gain $\text{G(t=1)}$ for (a) probe mode, $(\langle\hat{a}^{\dagger} \hat{a}\rangle_{t=1,\alpha = 0} - \langle\hat{a}^{\dagger} \hat{a}\rangle_{t=1,\alpha = 0.1})/\langle\hat{a}^{\dagger} \hat{a}\rangle_{t=1,\alpha = 0}$, and (b) conjugate mode, $(\langle\hat{b}^{\dagger} \hat{b}\rangle_{t=1,\alpha = 0} - \langle\hat{b}^{\dagger} \hat{b}\rangle_{t=1,\alpha = 0.1})/\langle\hat{b}^{\dagger} \hat{b}\rangle_{t=1,\alpha = 0}$. (c) The expectation value difference in percentile for the quantum correlation $|\langle \hat{a}\hat{b} \rangle|$ between $\alpha_a = 0.1$ and $\alpha_a = 0$, i.e., $(|\langle\hat{a}\hat{b}\rangle|_{t=1,\alpha = 0} - |\langle\hat{a}\hat{b}\rangle|_{t=1,\alpha = 0.1})/|\langle\hat{a}\hat{b}\rangle|_{t=1,\alpha = 0}$, versus the FWM gain.
    \label{Diff1}}
\end{figure*}

We denote $\hat{a}(t)$ and $\hat{b}(t)$ as the \textit{instantaneous} mode operators \textit{in the atomic $^{85}$Rb vapor cell} for the probe and conjugate modes respectively. From the equations of motion for these operators in the interaction picture,  one is able to obtain the time evolutions for operators $ \hat{a}$ and $\hat{b}^{\dagger}$:

\begin{equation}
\begin{aligned}
\partial_t{\hat{a}} = - ig \hat{b}^{\dagger} - \eta_a \hat{a} + f_a\hat{\nu}_a,\\
\partial_t{\hat{b}^{\dagger}} = ig \hat{a} - \eta_b\hat{b}^{\dagger} + f_b \hat{\nu}_b^{\dagger},
\label{eq1}
\end{aligned}
\end{equation}
where $g$ is related to the squeezing parameter $r$ by $r  \equiv gt$,  and $\eta_i$ ($i = a, b$) is related to the absorption parameter $\gamma_i$ by $\gamma_i  \equiv \eta_i t$. Then the \textit{instantaneous} gain $\text{G}(t)$ and absorption $\alpha_i(t)$ inside the cell at time $t$ can be written as

\begin{equation}
\begin{aligned}
\text{G}(t) = \text{cosh}^2 r = \text{cosh}^2 gt,\\
\alpha_i(t) = 1-e^{-2\gamma_i} = 1-e^{-2\eta_i t}. 
\label{eq2}
\end{aligned}
\end{equation}

The operators $\hat{\nu}_i$ ($i = a, b$) in Eq.~(\ref{eq1}) are the vacuum/noise operators and their corresponding prefactors $f_i$ are determined by obeying the commutation relations $[\hat{a}, \hat{a}^{\dagger}] =  [\hat{b}, \hat{b}^{\dagger}] = 1$. \\


Using Eq.~(\ref{eq1}) and the input state of $|0, 0, 0, 0\rangle$ (both probe and conjugate modes start from vacuum), time evolutions for the expectation values of $\langle\hat{a}^{\dagger} \hat{a}\rangle$, $\langle\hat{b}^{\dagger} \hat{b}\rangle$, $\langle\hat{a} \hat{b}\rangle$ and $\langle\hat{a}^{\dagger} \hat{b}^{\dagger} \rangle$ can be readily obtained:

\begin{equation}
\begin{split}
\partial_t{[\langle\hat{a}^{\dagger} \hat{a}\rangle]}&=\langle[\partial_t{\hat{a}}^{\dagger}] \cdot \hat{a}\rangle + \langle\hat{a}^{\dagger} \cdot [\partial_t{\hat{a}}]\rangle \\
 &= ig\langle\hat{a}\hat{b}\rangle - ig \langle \hat{a}^{\dagger} \hat{b}^{\dagger}\rangle - 2 \eta_a \langle \hat{a}^{\dagger} \hat{a}\rangle,
\label{eq3}
\end{split}
\end{equation}

\begin{equation}
\begin{split}
\partial_t{[\langle\hat{b}^{\dagger} \hat{b}\rangle]}&=\langle[\partial_t{\hat{b}}^{\dagger}] \cdot \hat{b}\rangle + \langle\hat{b}^{\dagger} \cdot [\partial_t{\hat{b}}]\rangle \\
&= ig\langle\hat{a}\hat{b}\rangle - ig \langle \hat{a}^{\dagger} \hat{b}^{\dagger}\rangle - 2 \eta_b \langle \hat{b}^{\dagger} \hat{b}\rangle,
\label{eq4}
\end{split}
\end{equation}

\begin{equation}
\begin{split}
\partial_t{[\langle\hat{a} \hat{b}\rangle]}&=\langle[\partial_t{\hat{a}}] \cdot \hat{b}\rangle + \langle\hat{a}\cdot [\partial_t{\hat{b}}]\rangle \\
&= - ig\langle\hat{b}^{\dagger}\hat{b}\rangle - ig (\langle \hat{a}^{\dagger} \hat{a}\rangle +1) -  (\eta_a+\eta_b) \langle \hat{a} \hat{b}\rangle,
\label{eq5}
\end{split}
\end{equation}

\begin{equation}
\begin{split}
\partial_t{[\langle\hat{a}^{\dagger} \hat{b}^{\dagger} \rangle]}&=\langle[\partial_t{\hat{a}}^{\dagger}] \cdot \hat{b}^{\dagger} \rangle + \langle\hat{a}^{\dagger} \cdot [\partial_t{\hat{b}^{\dagger}}]\rangle \\
&= ig\langle\hat{b}^{\dagger}\hat{b}\rangle + ig (\langle \hat{a}^{\dagger} \hat{a}\rangle +1) - (\eta_a+\eta_b) \langle \hat{a}^{\dagger}  \hat{b}^{\dagger} \rangle,
\label{eq6}
\end{split}
\end{equation}

Solving the above differential equations is straightforward with the initial conditions $\langle\hat{a}^{\dagger} \hat{a}\rangle_{t=0} = \langle\hat{b}^{\dagger} \hat{b}\rangle_{t=0} = \langle\hat{a} \hat{b}\rangle_{t=0} = \langle\hat{a}^{\dagger} \hat{b}^{\dagger} \rangle_{t=0} = 0$, i.e, the mean number of photons in the probe and conjugate modes and the mean quantum correlations are all zero when $t = 0$. \\

Following Eq.~(\ref{eq2}), the parameters $\eta_i$ and $g$ are given by $\eta_i = -[\text{ln}(1-\alpha_i)]/2t$ and $g = (\text{cosh}^{-1}\sqrt{\text{G}})/t$ respectively. For simplicity but without loss of generality, we set the total interaction time $t = 1$. We also assume that the total absorption on the probe mode through the cell is 0.1, i.e., $\alpha_a(t=1) = 0.1$ given that our maximal gain is only approximately 8, and the absorption on the conjugate mode is 0, i.e., $\alpha_b = \eta_b = 0$. This assumption is made according to Ref.~\cite{PhysRevA.78.043816}, where the simulation shows that when the gain is less than 10, the total loss on the probe mode due to atomic absorption of probe photons would be less than 10~\% (see Fig.~5 therein), and the absorption of conjugate photons would be negligible since it is far detuned from the atomic resonance.\\

In Fig.~\ref{Diff1}, we plot the mean number of photons difference in percentile at the exit of the cell between $\alpha_a = 0.1$ and $\alpha_a = 0$ versus the FWM gain $\text{G(t=1)}$ for the probe mode, $(\langle\hat{a}^{\dagger} \hat{a}\rangle_{t=1,\alpha = 0} - \langle\hat{a}^{\dagger} \hat{a}\rangle_{t=1,\alpha = 0.1})/\langle\hat{a}^{\dagger} \hat{a}\rangle_{t=1,\alpha = 0}$, in Fig.~\ref{Diff1}(a), and for the conjugate mode, $(\langle\hat{b}^{\dagger} \hat{b}\rangle_{t=1,\alpha = 0} - \langle\hat{b}^{\dagger} \hat{b}\rangle_{t=1,\alpha = 0.1})/\langle\hat{b}^{\dagger} \hat{b}\rangle_{t=1,\alpha = 0}$, in Fig.~\ref{Diff1}(b). The fact that there is merely at most 5~\% of reduction in the mean-photon-number generations with 10~\% absorption on the probe mode implies that the generated TMSV state is still very close to a pure state with a FWM gain of 8. It is therefore safe to assume that the data points in Fig.~\ref{ConevsPump} and Fig.~\ref{Decoh} are all in the pure states as their pump powers are less than 0.8~W, i.e., their FWM gains are less than 8.\\

In an effort to present a complete analysis, we also plot the expectation value difference in percentile for the quantum correlation $|\langle \hat{a}\hat{b} \rangle|$ between $\alpha_a = 0.1$ and $\alpha_a = 0$, i.e., $(|\langle\hat{a}\hat{b}\rangle|_{t=1,\alpha = 0} - |\langle\hat{a}\hat{b}\rangle|_{t=1,\alpha = 0.1})/|\langle\hat{a}\hat{b}\rangle|_{t=1,\alpha = 0}$, versus the FWM gain in Fig.~\ref{Diff1}(c). The less than 4~\% difference again indicates an inconsequential effect of the loss on the purity of the TMSV state with a FWM gain of less than 8. \\




\bibliography{MyLibrary}

\end{document}